\documentclass[a4paper,twocolumn,superscriptaddress,showpacs,preprintnumbers,amsmath,amssymb,prl]{revtex4}

\usepackage{graphicx}
\usepackage{dcolumn}
\usepackage{bm}
\usepackage[latin1]{inputenc}
\usepackage{epstopdf}
\usepackage{color}

\begin{document}

\preprint{Start}

\title{Splitting of the monolayer out-of-plane A$'_1$ Raman mode in few-layer WS$_2$}

\author{Matthias Staiger}
\email[Corresponding author. ]{mstaiger@physik.tu-berlin.de}
\affiliation{Institut für Festkörperphysik, Technische
Universit\"{a}t Berlin, Eugene-Wigner-Bldg. EW 5-4, Hardenbergstr.
36, 10623 Berlin, Germany}

\author{Roland Gillen}
\affiliation{Institut für Festkörperphysik, Technische
Universit\"{a}t Berlin, Eugene-Wigner-Bldg. EW 5-4, Hardenbergstr.
36, 10623 Berlin, Germany}

\author{Nils Scheuschner}
\affiliation{Institut für Festkörperphysik, Technische
Universit\"{a}t Berlin, Eugene-Wigner-Bldg. EW 5-4, Hardenbergstr.
36, 10623 Berlin, Germany}

\author{Oliver Ochedowski}
\affiliation{Fakultät für Physik and Cenide, Universität Duisburg-Essen, Lotharstr. 1, 47057 Duisburg, Germany}
\author{Felix Kampmann}
\affiliation{Institut für Festkörperphysik, Technische
Universit\"{a}t Berlin, Eugene-Wigner-Bldg. EW 5-4, Hardenbergstr.
36, 10623 Berlin, Germany}

\author{Marika Schleberger}
\affiliation{Fakultät für Physik and Cenide, Universität Duisburg-Essen, Lotharstr. 1, 47057 Duisburg, Germany}

\author{Christian Thomsen}
\affiliation{Institut für Festkörperphysik, Technische
Universit\"{a}t Berlin, Eugene-Wigner-Bldg. EW 5-4, Hardenbergstr.
36, 10623 Berlin, Germany}

\author{Janina Maultzsch}
\affiliation{Institut für Festkörperphysik, Technische
Universit\"{a}t Berlin, Eugene-Wigner-Bldg. EW 5-4, Hardenbergstr.
36, 10623 Berlin, Germany}

\begin{abstract}

We present Raman measurements of mono- and few-layer WS$_2$. We study the monolayer $A'_1$ mode around 420\,cm$^{-1}$ and its evolution with the number of layers.  We show that with increasing layer number there is an increasing number of possible vibrational patterns for the out-of-plane Raman mode: in \textit{N}-layer WS$_2$ there are $N$ $\Gamma$-point phonons evolving from the $A'_1$ monolayer mode. For an excitation energy close to resonance with the $A$ excitonic transition energy, we were able to observe all of these $N$ components, irrespective of their Raman activity. Density functional theory calculations support the experimental findings and make it possible to attribute the modes to their respective symmetries. The findings described here are of general importance for all other phonon modes in WS$_2$ and other layered transition metal dichalcogenide systems in the few layer regime. 


\end{abstract}

\maketitle

\section*{Introduction} 

The last few years have seen a spectacular increase in interest in layered transition metal dichalcogenides (TMDs). 
While the properties of the 3D bulk materials have been well known for decades, the possibility of thinning them down towards the monolayer has given rise to an entirely new research area.  
Their structural formula $MX_2$ (with $M$ being a transition metal and $X$ a chalcogenide) comprises metals, semimetals, semiconductors and superconductors \cite{Chhowalla2013}.
What they have in common is that, with approaching the 2D limit, a whole world of intriguing properties such as extraordinarily large exciton binding energies \cite{Ye2014}, robust valley polarization \cite{Zhu2014} and large spin orbit splitting \cite{Ramasubramaniam2012} opens up.
These properties pave the way for potential use of TMDs for applications in digital electronics and optoelectronics \cite{Wang2012}, in energy conversion and storage \cite{Wang2014} as well as in spintronics \cite{Zhu2014a}. 
Most tungsten and molybdenum based TMDs exhibit a transition from an indirect to a direct bandgap semiconductor when being thinned down to the monolayer, resulting in high intensity photoluminescence \cite{Splendiani2010,Mak2010,Tonndorf2013}.  \\
Raman spectroscopy is one of the most powerful tools in characterizing nanomaterials. For few-layer (FL-) TMDs  such as FL-WS$_2$ it allows for instance to exactly determine the number of layers. To be able to extract all the information that the measured Raman spectra have to offer, it is of primary importance to have a common ground on which to categorize the different Raman modes as a function of the number of layers. 
There is a large number of publications on monolayer TMDs \cite{Matte2010,Molina2011,Thripuranthaka2014,Mitioglu2014} and the respective bulk materials \cite{Wieting1971,Sourisseau1989,Molina2011}, but only few articles concentrate on the transition from monolayer to bulk, e.g. the evolvement of the Raman signatures with the number of layers. In these articles moreover, many use the symmetry of the bulk to assign the Raman features \cite{Lee2010,Plechinger2012,Berkdemir2013,Tonndorf2013}; only recently there have been some reports that take into account the different symmetries of even number of layers (even \textit{N}) and odd number of layers (odd \textit{N}) TMDs \cite{Zhang2013,Terrones2014}. However they only focus on Raman modes that are allowed in first-order scattering. \\
In this work we will show that (i) it is important to distinguish between even and odd number of layers and (ii) that when the excitation energy is in resonance with the first optical transition of the investigated material it is necessary to consider the full set of phonons. We study the splitting of the monolayer out-of-plane A$'_1$ mode in FL-WS$_2$ in particular and are able to observe layer dependent Raman signatures comprising Raman active and inactive modes. They allow for an easy identification of the number of layers via Raman spectroscopy. Moreover a general systematic behavior for the splitting of the monolayer Raman modes is proven experimentally for the first time and supported by DFT calculations.  The findings described in this work should be expandable to other FL-TMDs.

\begin{figure}[t!]
\includegraphics[width=0.5\textwidth]{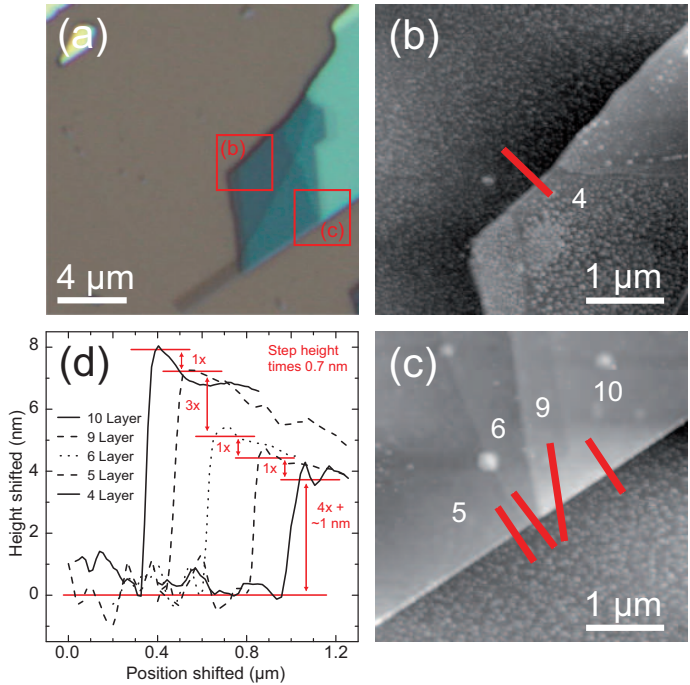}
\caption{(a) Optical micrograph of a WS$_2$ flake with selected regions inspected by AFM [shown in (b) and (c)].(b) and (c) AFM images of the selected regions with indication of layer numbers and lines showing where AFM height profiles were taken from. (d) AFM height profiles going from the substrate onto the flakes for different positions on the flake. Step heights in units of the observed interplanar spacing of $\sim$0.7\,nm are indicated.}
\end{figure}

\begin{figure*}[t]
\includegraphics[width=1.0\textwidth]{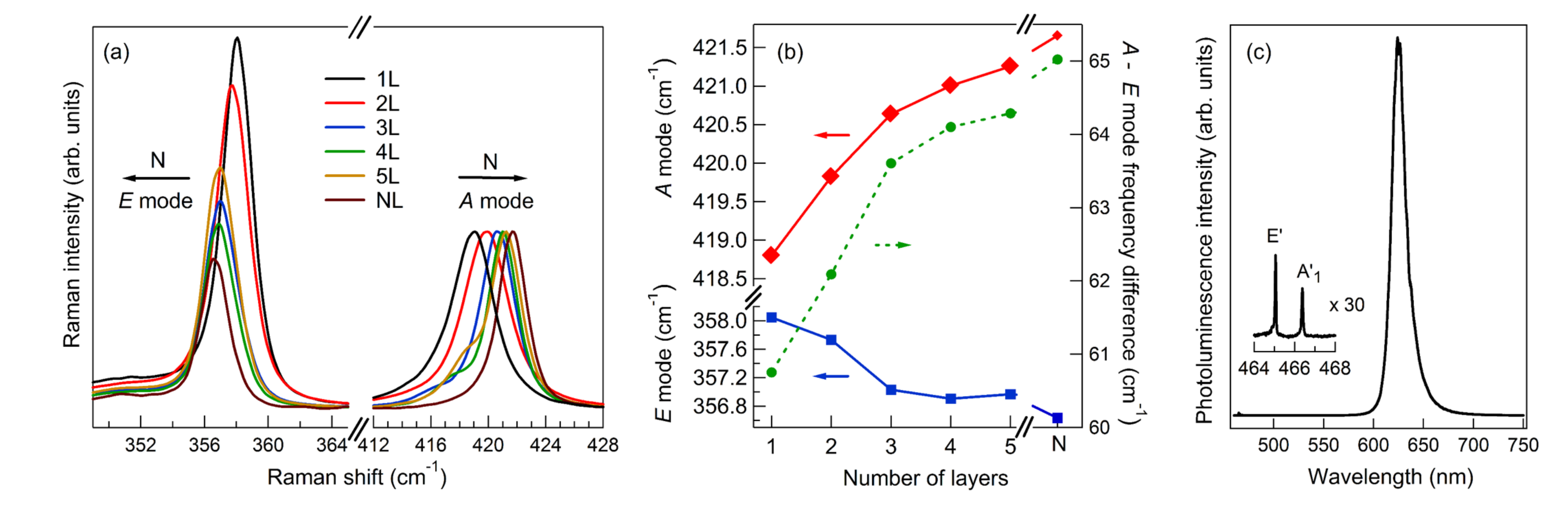}
\caption{(a) Raman spectra of FL WS$_2$ from one (1L) to five (5L) layers and the bulk material (NL). Excitation wavelength is 457\,nm. Spectra are normalized to the intensity of the \textit{A} mode at 420\,cm$^{-1}$. (b) Evolution of the frequency of the main \textit{E} and \textit{A} mode with the number of layers. The dashed line marks the increasing frequency difference between the modes with the number of layers. (c) Strong photoluminescence is observed for monolayer WS$_2$ at the direct excitonic transition energy around 625\,nm. The inset shows a magnified image of the same spectrum in the region of the main 1L-WS$_2$ Raman peaks.}
\end{figure*}

\section*{Experimental}
The samples are prepared from a bulk WS$_2$ crystal (hq graphene, Netherlands - Groningen) using the mechanical exfoliation technique. To enhance the optical contrast, the crystals are exfoliated onto a 90\,nm SiO$_2$/Si wafer. 
Raman measurements on WS$_2$ samples were done at room temperature in backscattering geometry with a Horiba Jobin Yvon LabRAM HR spectrometer using a confocal setup with a 100x objective and excitation wavelengths of 457\,nm  and 633\,nm. To avoid sample heating the laser power was kept below a maximum of 120\,$\mu$W; a 1800 lines mm$^{-1}$ grid was used to ensure high spectral resolution of around 1\,cm$^{-1}$. First, Raman spectra were taken with the setup described above and calibrated with Neon lines. In a second step the measurements were repeated in subpixel (6sp) mode. There, each spectrum is taken a couple of times, while each time the spectrometer is shifted by a step size which is smaller than a pixel value. This technique does not increase the spectral resolution, however, by providing more data points per wavenumber, it reduces the signal to noise ratio significantly. The spectra aquired in subpixel mode were then shifted in frequency to match the calibrated Raman spectra obtained in the regular single-window mode. For better comparison, in Figs. 2, 3, 4, the spectra were normalized to the intensity of the out-of-plane mode. \\
Atomic force microscopy (AFM) images were acquired using a Park Systems XE-100 setup with commercial silicon tips in tapping mode configuration. Images were taken with 256x256\,px resolution (4x4\,$\mu$m). An exemplary AFM analysis of a FL-WS$_2$ sample is shown in Fig. 1. Step heights between subsequent layer numbers were typically around 0.7\,nm, close to the experimental value of the interplanar spacing of WS$_2$ layers \cite{Wilson1969}. Down to the monolayer, an offset of typically around 1\,nm between substrate and sample was observed, which is probably due to the presence of adsorbates in between the substrate and the sample and was taken into account in the analysis presented in Fig. 1. \\ The phonon frequencies of monolayer and FL-WS$_2$ at the $\Gamma $-point were calculated in the frame of density functional (perturbation) theory on the level of the local density approximation (LDA) as implemented in the CASTEP code \cite{Clark2005}. We treated the W(4d,5s) and the S(3s,3p) states as valence electrons using normconserving pseudopotentials with a cutoff energy of 800\,eV. All reciprocal space integrations were performed by a discrete $k$-point sampling of 18x18x1 $k$-points in the Brillouin zone. Starting from fully symmetric model geometries of one to five layers of AB-stacked WS$_2$, we fully optimized the lattice constants and atomic positions until the residual forces between atoms were smaller than 0.01\,eV/\r{A} and the stresses on the cell boundaries were smaller than 2.5$\times10^{-3}$\,GPa. The obtained in-plane lattice constants slightly increased with layer number from a value of 3.141\,\r{A}  for 1L-WS$_2$ to 3.143\,\r{A} for 5L-WS$_2$, in good agreement with the experimental in-plane lattice constant of 3.15\,\r{A} in bulk WS$_2$ \cite{Schutte1987}. Interactions of the sheet with residual periodic images due to the 3D boundary conditions were minimized by maintaining a vacuum layer of at least 20\,\r{A}. 

\section*{Results}

WS$_2$, like MoS$_2$, crystallizes in the 2H trigonal prismatic structure where the tungsten atoms are sandwiched between two layers of sulfur atoms. Intralayer bonds are of covalent nature, the interlayer interaction is governed by weak van-der-Waals forces.
Bulk WS$_2$ has the D$_{6h}$ point group; the 6 atoms per unit cell result in 18 phonon modes at the $\Gamma$-point of the hexagonal Brillouin zone \cite{Verble1970}:
\\[0.2cm]
$D_{6h} : \Gamma=A_{1g}+2A_{2u}+B_{1u}+2B_{2g}+E_{1g}+2E_{1u}+E_{2u}+2E_{2g}$
\\[0.2cm]
\noindent In the monolayer and an odd number of layers (odd \textit{N}), the symmetry is reduced to the D$_{3h}$ point group. Therefore, odd \textit{N} WS$_2$ do not have a center of inversion. The $\Gamma$-point phonon modes transform according to the following irreducible representation: 
\\[0.2cm]
$D_{3h} : \Gamma=\frac{3\textit{N}-1}{2}(A'_1+A''_2+E'+E'')+A''_2+E',\\ N=1,3,5,...$
\\[0.2cm]
\noindent For even number of layers (even \textit{N}), WS$_2$ possesses a center of inversion, the symmetry is described by the point group D$_{3d}$: 
\\[0.2cm]
$D_{3d} : \Gamma=\frac{3\textit{N}}{2}(A_{1g}+E_g+A_{2u}+E_u), \\N=2,4,6,...$
\\[0.2cm]
\noindent Let us now first consider the case where the excitation wavelength is far from resonance. Figure 2 (a) shows Raman spectra of one to five WS$_2$ layers and the bulk material taken with an excitation wavelength of 457\,nm. The two main Raman modes are the A$'_1$  and A$_{1g}$ mode around 420\,cm$^{-1}$ for odd and even \textit{N}, respectively, and the E$'$ and E$^1_{2g}$ mode around 355\,cm$^{-1}$ for odd \textit{N} and even \textit{N}, respectively. A clear upshift with increasing number of layers is seen for the out-of-plane A$'_1$/A$_{1g}$ mode, whereas the in-plane E$'$/E$^1_{2g}$ mode slightly softens. This has been observed before for WS$_2$ \cite{Zhao2013,Berkdemir2013} as well as for other TMDs \cite{Lee2010,Late2012,Tonndorf2013,Yamamoto2014}. The stiffening of the out-of-plane \textit{A} mode is explained by the increasing interlayer interaction and the subsequent rise in restoring forces on the atoms with the number of layers \cite{Lee2010}. The same should hold for the \textit{E} mode, although to a lesser extent, as the atoms move in-plane and the influence of interlayer interaction is thus expected to be smaller. However, the opposite trend is observed and has been attributed to dielectric screening of long range Coulomb interactions \cite{Molina2011}. Figure 2 (b) depicts the change in frequency of the out-of-plane \textit{A} and in-plane \textit{E} mode with the number of layers and the frequency difference between the two modes, which increases from 60.4\,cm$^{-1}$ to 65\,cm$^{-1}$ from the monolayer to the bulk.\\
If the exciting light is far from resonance, the spectra are dominated by Raman modes allowed in first-order scattering, as has been reported previously for WS$_2$ nanotubes \cite{Staiger2012}. If we focus on the out-of-plane \textit{A} modes, it is evident that the mono-, bilayer and bulk spectra show a single peak, whereas for three and more layers there is at least another mode appearing as a low-energy shoulder of the dominant Raman feature. This apparent splitting of the out-of-plane mode is due to the fact that for monolayer and bulk WS$_2$ there is only one Raman active A$'_1$ and A$_{1g}$ mode, respectively; for few layers starting with three layers, more than one Raman mode becomes allowed, see also Ref. \cite{Terrones2014}.\\
To further explore these new Raman modes, which are reported here for the first time in FL-WS$_2$, we analyze the Raman spectra of the same samples taken under the resonance condition. Figure 2 (c) shows a photoluminescence spectrum of monolayer WS$_2$ at 457\,nm excitation wavelength. The photoluminescence signal is more than two orders of magnitude larger than the Raman signal and has maximum intensity around 625\,nm. From previous experiments it is known that the first optical transition energy of bulk WS$_2$ is constituted by the $A$ exciton around 633\,nm \cite{Staiger2012}. Therefore, with 633\,nm excitation wavelength, we are close to the $A$ excitonic resonance for mono- and few layer WS$_2$. In the remainder of this work we will focus on the out-of-plane \textit{A} mode around 420\,cm$^{-1}$. Of the four Raman modes allowed in bulk WS$_2$ D$_{6h}$ symmetry, the E$^2_{2g}$ is too low in frequency to be observed here, the E$_{1g}$ mode is not allowed in backscattering geometry, and the E$^1_{2g}$ mode around 350\,cm$^{-1}$ overlaps with a second order mode, which dominates the spectra in resonance \cite{Sourisseau1989}. As we will show below, the out-of-plane mode is clearly separated from other Raman features and shows sidebands that can readily be explained by the respective symmetries of even or odd number of layers. 

\begin{figure}[t]
\includegraphics[width=0.5\textwidth]{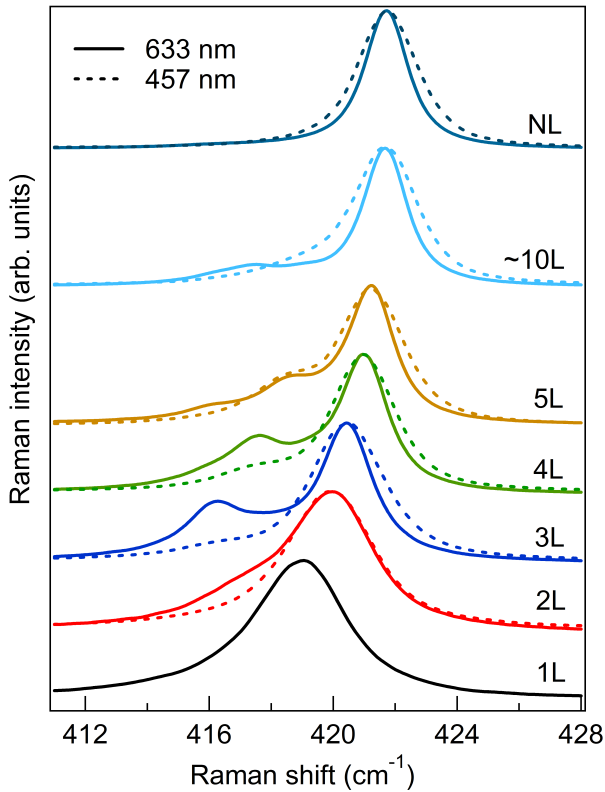}
\caption{Experimental resonance Raman spectra (excitation wavelength 633\,nm) of the \textit{A} mode region in FL-WS$_2$ from the bilayer (2L) to five layers (5L) and the bulk (NL).  The monolayer spectrum (1L) is only shown with an excitation wavelength of 457\,nm, as the Raman features are dominated by strong photoluminescence at 633\,nm excitation spectrum (not shown). For FL-WS$_2$, spectra of the same region taken with 457\,nm excitation wavelength are overlayed and depicted with dashed lines. Spectra are normalized to the main Raman peak and are offset for clarity.}
\end{figure}

\begin{figure*}[t]
\includegraphics[width=1.0\textwidth]{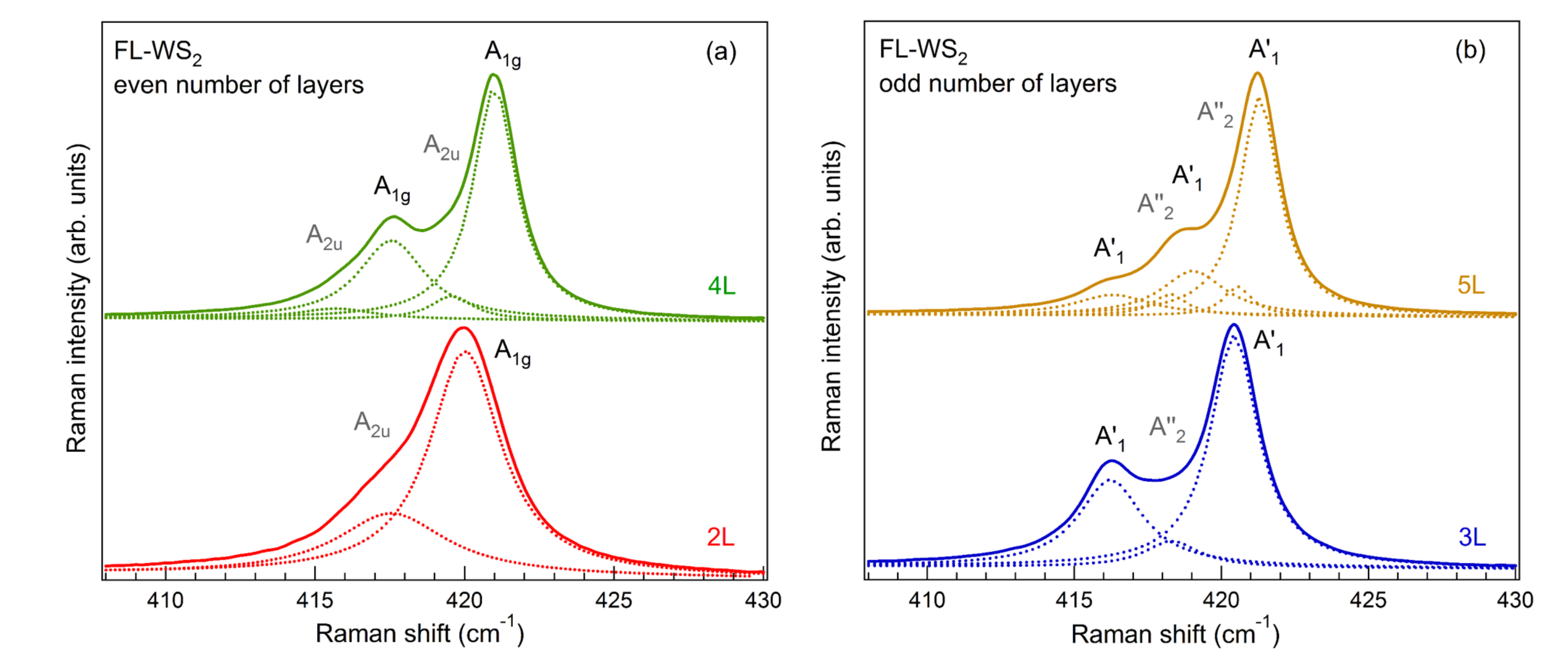}
\caption{Experimental resonance Raman spectra of FL-WS$_2$. The Lorentzian fit curves are shown as well as the symmetry attributed to the individual modes. In general, there are \textit{N} components for a \textit{N} layer spectrum. (a) For even number of layers, Raman active A$_{1g}$ modes alternate with infrared active A$_{2u}$ modes. (b) For odd number of layers, Raman active A$'_1$ modes alternate with infrared A$''_2$ modes. For both, even and odd \textit{N}, the Raman active modes are more intense than the rather weak infrared active modes.}
\end{figure*}

Figure 3 shows the region of the out-of-plane Raman mode of few-layer WS$_2$ taken with 633\,nm excitation wavelength. Similarly to the spectra shown at 457\,nm excitation wavelength, the upshift of the Raman mode with the number of layers is evident. More importantly though, there are striking differences in the shape of the Raman mode. For FL-WS$_2$ the structure of the out-of-plane mode is complex. In contrast to the spectra described above, already the bilayer spectrum shows more than one component. More and more sidebands arise for increasing number of layers but they get weaker for $\textrm{n}>5$ and vanish for very thick flakes ($\textit{n}=N$). This again underlines the special role played by few-layer samples: for $\textit{n}=1$ and $\textit{n}=N$ the symmetry of WS$_2$ allows only one A$'_1$ ($\textit{n}=1$) and A$_{1g}$ (bulk) Raman mode. The comparison with the out-of-resonance spectra (457\,nm excitation wavelength) illustrates that, even though shoulders of the main Raman peak are observed at 457\,nm excitation as well, an increased number of well pronounced sidebands to the main Raman peak appears mainly for Raman measurements in resonance with the $A$ exciton. 
We fit the spectra with Lorentzian profiles, see Fig. 4; for clarity spectra of even and odd \textit{N} are shown in separate graphs. In Fig. 4 (a) the bilayer A$_{1g}$ mode spectrum possesses a low energy shoulder that has not previously been observed. Its appearance is surprising since the only expected Raman active vibration is the A$_{1g}$ mode, where both layers vibrate in-phase according to the monolayer A$'_1$ mode. We will show below that this second peak is indeed the infrared active A$_{2u}$ mode, where the two layers vibrate out-of-phase. In the four-layer (4L) spectrum the dominant A$_{1g}$ mode shifts up with respect to the bilayer, and a prominent shoulder is seen at almost the same frequency as the shoulder in the bilayer spectrum. The 4L spectrum is best fitted with four Lorentzians to account for the Raman intensity between the two stronger Raman features. In the spectra of odd \textit{N} WS$_2$, a similar pattern is revealed. In Fig. 4 (b) the trilayer (3L) and five-layer (5L) spectra are shown. The former consists of the expected A$'_1$ peak as the most significant contribution and a pronounced low-energy shoulder. A third Lorentzian fits the plateau between the main peaks.  The 3L spectrum cannot be properly fitted with only two Lorentzians. From the experience gained from the spectra investigated above, the five-layer spectrum is fitted with five Lorentzians, two of which fill up the region between the main A$'_1$ peak and the two low-energy shoulders. From the spectra for layer numbers of $\textit{n}=1$  to $\textit{n}=5$ it thus seems that there are always \textit{N} components to the out-of-plane \textit{A} mode, where \textit{N} is the number of layers. We have observed sidebands on the lower energy side of the dominant Raman peak also for higher layer numbers as exemplary shown for the spectrum of an around 10 layer thick flake. But they are rather weak and cannot be fitted following the pattern described above. 

\begin{figure}[t]
\includegraphics[width=0.28\textwidth]{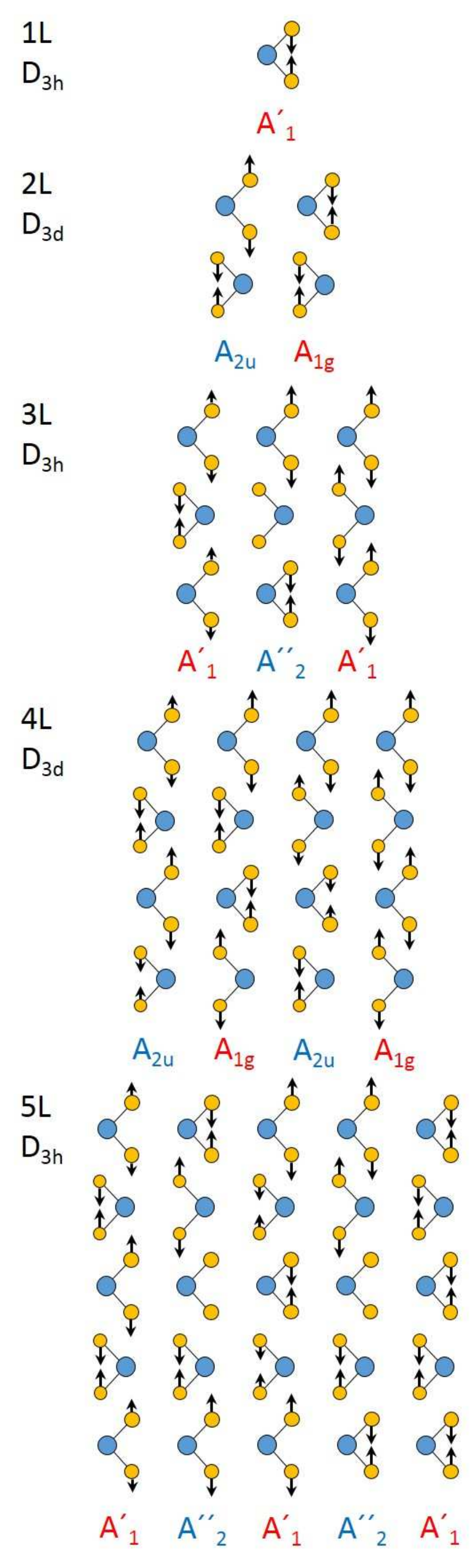}
\caption{Schematic drawing for all possible vibrational modes in the out-of-plane mode region of FL-WS$_2$ (Raman and infrared) for one (1L) to five (5L) layers together with the symmetry assignments taken from our DFT calculations. The displacement patterns are ordered with increasing frequency from left to right. Taking into account the full set of possible vibrational patterns helps to study the splitting of vibrational modes with increasing layer number and to attribute the modes to the features seen in the Raman spectra.}
\end{figure}

\section{Discussion}

Recently, careful analysis of few-layer TMDs has led to the observation of Raman modes that are neither seen in the bulk nor in the monolayer \cite{Tonndorf2013}. Some of them appear as shoulders to Raman modes that are allowed in first order for bulk and monolayer, like the bulk A$_{1g}$ mode discussed here. Others are Raman inactive or not allowed in backscattering geometry in bulk and monolayer, like the bulk B$_{2g}$ and E$_{1g}$ modes \cite{Luo2013,Terrones2014,Tonndorf2013,Scheuschner2015}.
For the first case, to the best of our knowledge, there is only one article that explicitly shows a splitting of the first-order A$_{1g}$ mode with the number of layers in few-layer MoSe$_2$ \cite{Tonndorf2013}. The 3L and 4L samples show two components, for the 5L sample a third component is seen. For WSe$_2$ the overlap of the bulk E$^1_{2g}$ and A$_{1g}$ mode makes an observation of such shoulders impossible \cite{Luo2013,Terrones2014}, and for the case of MoTe$_2$ the intensity of the A$_{1g}$ mode appears to be too small to resolve a multipeak structure \cite{Yamamoto2014}. For MoS$_2$ there is only little literature on resonance Raman spectra, and despite some asymmetry in the shape of the out-of-plane Raman mode, a splitting similar to the one investigated in this work, is not observed \cite{Scheuschner2012,Li2012a}. Terrones et al. \cite{Terrones2014} calculate the optical phonons for a number of FL-TMDs, among them MoS$_2$ and WS$_2$, but do not discuss phonons other than the Raman active ones.\\
In order to have a theoretical background to the experimentally observed appearance of more than just the Raman active vibrational modes in the spectra of FL-WS$_2$, we have done calculations employing density functional theory (DFT). A better insight into the atomic displacements corresponding to the phonon modes of FL-TMDs is given in Fig. 5, where schematic drawings of all possible vibrations evolving from the monolayer A$'_1$ mode in WS$_2$ from $\textit{n}=1-5$ are depicted, based on the DFT calculations. While there is only one possibility in monolayer WS$_2$ for the sulfur atoms to vibrate against each other with a fixed tungsten atom in between, a splitting of this mode occurs for bilayer WS$_2$. Since bilayer WS$_2$ possesses a center of inversion, there is a Raman active A$_{1g}$ mode, where the two layers vibrate in phase, and an infrared active A$_{2u}$ mode, where the two layers vibrate out of phase. As the latter is not Raman active it is not seen in Raman spectra taken with excitation wavelengths far from resonance [Fig. 2 (a)]. However, it is observed for the resonance Raman spectrum (Figs. 3 and 4), albeit with weaker intensity than the dominant A$_{1g}$ mode. Several possible reasons for this unusual behavior are discussed below. As the two layers interact more strongly for the in-phase vibration, the A$_{1g}$ mode has a slightly higher frequency than its infrared active counterpart. For 4L-WS$_2$, each of the two bilayer modes again splits up into a Raman-active A$_{1g}$ and a infrared active A$_{2u}$ mode. The spectrum is still governed by the in-phase vibration of all four layers (A$_{1g}$) but there is a second Raman active A$_{1g}$ mode that has the outer layers vibrating out of phase with the inner ones, thus retaining the inversion symmetry of the overall structure. It is interesting to note that this lower lying A$_{1g}$ mode in the 4L spectrum has almost the same frequency as the infrared active mode in the bilayer, a pattern that will also be observed for odd \textit{N}, see below. In addition, we identify the two small shoulders on the lower-frequency side of the two Raman active modes with the A$_{2u}$ modes [Fig. 4 (a)], the lowest lying with neighboring layers vibrating out of phase, the other one with the two upper layers vibrating out of phase with the two lower layers.\\
In odd \textit{N} WS$_2$ there is obviously again the possibility of all layers vibrating in-phase and out-of-phase. In contrast to even \textit{N} WS$_2$, where the pure out of phase vibration is not Raman active, for odd \textit{N} both, in- and out-of-phase vibrations, are Raman active and possess A$'_1$ symmetry. For trilayer WS$_2$, the atomic displacement vectors of these two modes are shown on the right and left side of the third panel of Fig. 5.  In the spectrum depicted in Fig. 4 (b), the lower lying A$'_1$ mode accounts for the strong shoulder at approximately 416\,cm$^{-1}$ of the main A$'_1$ mode (in-phase vibration). In between the modes a plateau is evident that is not accounted for if the spectrum is only fitted with two Lorentzians. The origin of the plateau is attributed to an infrared active A$''_2$ mode, with the middle layer fixed and the sulfur atoms of the top and bottom layer vibrating out-of-phase (see Fig. 5, third panel, middle). The same approach can now be used for the analysis of the five layer WS$_2$ spectrum. Two shoulders to the main A$'_1$ peak can be identified and, following the pattern of Fig. 5, attributed to another two Raman active A$'_1$ modes. In between the Raman active modes two very weak features belong to infrared active A$''_2$ vibrations. Coming from the trilayer WS$_2$, the two modes with the highest frequency in the five-layer material can be imagined as stemming from a splitting of the main Raman active A$'_1$ mode in trilayer. The same can be said about the lower frequency Raman active mode in the trilayer that splits up into the low frequency A$'_1$ and A$''_2$ mode in five layers. The only infrared active A$''_2$ mode in the trilayer spectrum changes symmetry in the five layer and has the two outermost layers vibrating out-of-phase with the three inner ones.\\ 
To summarize, for analyzing the Raman spectra of one to five layers of WS$_2$, it is necessary to take into account the full set of phonon modes. For the evolvement of the monolayer A$'_1$ mode analysed in detail here, the number of out-of-plane modes matches the number of layers. Raman active modes alternate with modes that are infrared active and are seen in the spectra of FL-WS$_2$ for the first time. 
The pattern observed for the out-of-plane mode should also be valid for all other vibrational modes in FL-WS$_2$ and other FL-TMD materials. For even \textit{N}, the monolayer mode splits up into $\frac{\textit{N}}{2}$ Raman active and $\frac{\textit{N}}{2}$ infrared active modes (the in-plane modes are always doubly degenerate). For odd \textit{N}, the monolayer mode evolves into $\frac{\textit{N}+1}{2}$ Raman active and $\frac{\textit{N}-1}{2}$ infrared active components. In total, in \textit{N} layers, each of the monolayer phonons splits up into \textit{N} phonon modes \cite{Scheuschner2015,Luo2013}.\\

\begin{table}
\begin{ruledtabular}
\begin{tabular}{llllll}
  &        1L&  2L       &3L &4L &5L \\
\hline
R&A$'_1$(418.8)& A$_{1g}$ (420.0) & A$'_1$ (420.5)       & A$_{1g}$ (421.0)  & A$'_1$ (421.3)   \\
 IR&          & A$_{2u}$ (417.3) & A$''_2$ (418.5)      & A$_{2u}$ (419.9)  & A$''_2$ (420.4)  \\
 R&  &                         & A$'_1$ (416.2)        &A$_{1g}$ (417.8)   & A$'_1$ (418.8)  \\
IR&  &  &                                         & A$_{2u}$ (416.0)   & A$''_2$ (418.0)              \\
R&  &  &  &                                                          & A$'_1$ (416.2)     \\
\end{tabular}
\end{ruledtabular}
\caption{Experimental Raman frequencies (in cm$^{-1}$) of all vibrational modes observed in FL-WS$_2$ excited with 633\,nm. For \textit{N} layers the A$'_1$ mode of the monolayer splits up into \textit{N} components. The point group of FL-WS$_2$ is D$_{3h}$ for odd number of layers and D$_{3d}$ for even number of layers. The vibrational modes for every number of layers alternate between Raman and infrared active. \label{tab1}}
\end{table}

\begin{figure*}[t]
\includegraphics[width=1.0\textwidth]{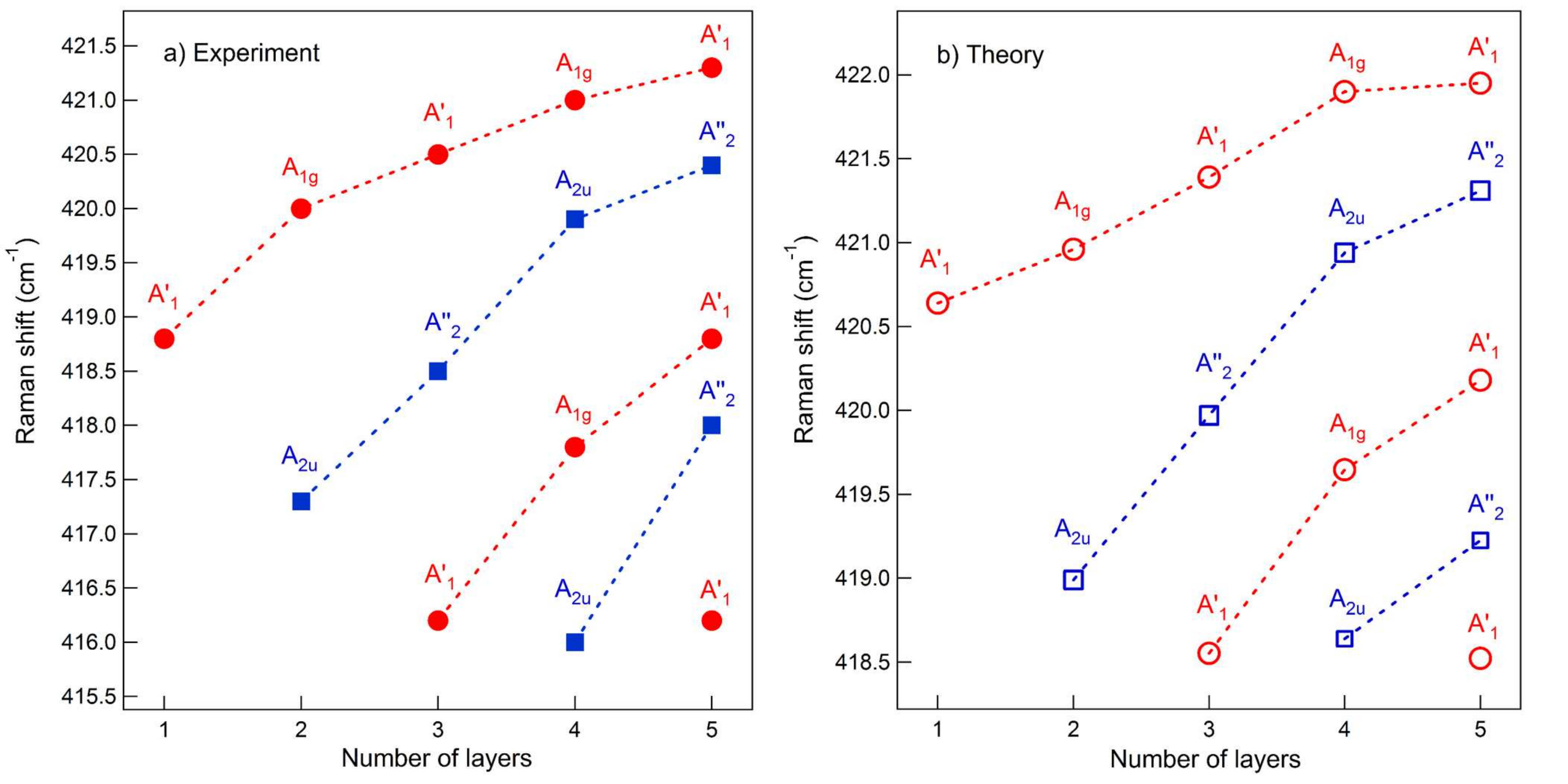}
\caption{a) Raman frequencies of the \textit{A} type modes in FL-WS$_2$. Red circles mark the Raman active modes, blue squares denote the infrared active modes also seen in the Raman spectra. Dashed lines are used to guide the eye. Starting from one layer the splitting of the modes for FL-WS$_2$ produces a fan-like shape of the possible vibrational frequencies. b) Theoretical calculations with DFT reproduce the observed trends. Raman active modes are denoted with open red circles, infrared active modes are marked with open blue squares.}
\end{figure*}

Table 1 lists the experimentally obtained Raman frequencies of all out-of-plane vibrations derived from the A$'_1$ mode at 418.8\,cm$^{-1}$  for few-layer WS$_2$. Where two or more samples with the same number of layers were measured, the average value is given in Table 1. In all these cases deviations from the given frequencies are less than 0.3\,cm$^{-1}$. Again, the table illustrates that Raman and infrared active modes are alternating irrespective of the layer number. Additionally, not only the main Raman active component with A$'_1$/A$_{1g}$ symmetry for odd and even \textit{N} appears to exhibit increased frequency with increasing number of layers but also the other components (2nd row and below in Table 1) follow the same trend. This is illustrated in Fig. 6 (a), where the tabulated frequencies are plotted against the number of layers. The main Raman peak in the spectra shown in Figs. 2-4 is seen stiffening in frequency from the monolayer to five layers (red circles, connected by a dashed line to guide the eye). Starting with the bilayer an infrared active mode comes into play that also appears for higher layer numbers. It also increases in frequency, thus following the behavior of the main Raman component due to increased force constants with increasing number of layers (blue squares, connected by a dashed line to guide the eye). The same is seen for a second Raman active feature starting with three layers and another infrared active mode starting with four layers. Interestingly, the frequencies observed in the mono- , bi- and trilayer are almost exactly repeated when the layer number is increased by two, underlining the close relation of the out-of-plane modes even though the symmetry and Raman/infrared activity changes with the layer number. In contrast, the position of the lowest frequency mode stays almost constant from three layer onwards. This mode always has neighboring layers moving out-of-phase, but neighboring sulfur atoms from adjacent layers moving in-phase. As a result, the nearest neighbor force constants determining the frequency of this mode will not change significantly for larger number of layers.
In Fig. 6 (b), the frequencies calculated with DFT are plotted against the layer number. Despite a slight overestimation of absolute frequencies and a smaller magnitude of the splitting of the modes, the experimental results are well reproduced. As a whole, the splitting of the out-of-plane monolayer A$'_1$ mode in FL-WS$_2$ results in a fan-like shape showing similarity to Fig. 5 in Ref. \cite{Zhang2013}. Zhang \textit{et al.} \cite{Zhang2013} investigated the evolvement of the low-frequency rigid layer C (displacement along the \textit{c}-direction) and LB (layer breathing) modes in FL-MoS$_2$, a first report on these interlayer
modes in few-layer MoS$_2$ can be found in Ref. \cite{Zeng2012}). With the support of a simple atomic chain model the authors found a behavior similar to the one described here for the out-of-plane vibration in WS$_2$. Obviously, rigid layer vibrations only appear starting from the bilayer. Strictly speaking, the difference to all other optical modes in FL-TMDs is that there are consequently only \textit{N}-1 possible vibrations (where \textit{N} denotes the layer number) for the rigid layer modes (E$^2_{2g}$ and B$_{2g}$ symmetry in bulk MoS$_2$). Of course, the ``\textit{N} modes for \textit{N} layer rule'' is restored again if one adds the acoustic modes. The acoustic E$'$ and A$''_2$ modes of the monolayer are the origin of the low frequency vibrations in FL-TMDs and will split up into \textit{N} components for \textit{N} layers (among them an \textit{A} and an \textit{E} type mode with zero frequency).\\
The behavior described above should in principle be observable in other 2H-TMDs as well. Terrones \textit{et al.} \cite{Terrones2014} predict an increased splitting of the Raman active components in the order WSe$_2$, MoSe$_2$, WS$_2$ and MoS$_2$. What distinguishes WS$_2$ from other prominent 2H-TMDs like WSe$_2$ and MoS$_2$ is that the bulk A$_{1g}$ and E$^1_{2g}$ modes are well separated in energy and that no second-order Raman features overlap with the A$_{1g}$ mode, thus making it easier to resolve the splitting of the out-of-plane mode into Raman and infrared active components. More importantly, measuring in resonance with the optical transition appears to be a necessary condition to observe the full set of vibrational modes. There is a lack of studies on Raman spectra of other FL-2H-TMDs measured under resonance condition; often the corresponding excitation wavelengths are avoided because in the monolayer case the Raman features are obscured by strong photoluminescence signal. In FL-WS$_2$ in particular, the characteristic shape of the out-of-plane mode in resonance Raman spectra can be used as a fingerprint region to unambiguously identify the number of layers.\\
For WS$_2$ measured under the resonance excitation, even in the bulk material, the A$_{1g}$ out-of-plane mode is accompanied by a small shoulder that is attributed to the silent B$_{1u}$ mode \cite{Sourisseau1989,Staiger2012,Molina2011}. It had been previously attributed to a \textit{LA(K)+TA(K)} combination mode \cite{Sourisseau1991}, but the participation of two phonon modes involving acoustic phonons in this frequency region seems unlikely, especially in light of a more recent DFT calculations showing that nowhere in the Brillouin zone the acousting phonons reach values above 200\,cm$^{-1}$ \cite{Molina2011}. The in-phase and out-of-phase vibrational pattern of the A$_{1g}$/B$_{1u}$ pair in bulk WS$_2$ finds its counterpart in the variety of in- and out-of-phase vibrations observed in FL-WS$_2$. In WS$_2$ nanomaterials, the B$_{1u}$ mode gains in Raman intensity and its evolvement can be followed in WS$_2$ nanomaterials under pressure \cite{Yu2007,Staiger2012}, in different layer orientation in thin films \cite{Chung1998} and in different diameter WS$_2$ nanotubes \cite{Krause2009,Staiger2012}. In these cases, a strong resonance behavior is observed as well, much like in FL-WS$_2$: the modes not allowed in a first-order Raman process appear most strongly when the excitation energy is in or close to resonance with the optical transitions. 
In an earlier work on WS$_2$ nanotubes \cite{Staiger2012}, we found that the curvature-induced strain and the resultant crystal symmetry distortion was to be held accountable for the activation of silent modes. Here, in quasi 2D materials, the situation in absence of curvature effects is different; strain due to substrate-sample interaction is assumed to only play a very minor role in Raman spectroscopy on supported FL-TMDs \cite{Zhang2013}.\\
Instead, a closer look at the nature of the excitonic transition leading to the resonantly enhanced Raman intensity in the spectra of FL-WS$_2$ can provide means to elucidate the appearance of the infrared-active components of the A$'_1$/A$_{1g}$ Raman mode around 420\,cm$^{-1}$. For 633\,nm excitation wavelength, the phonons are coupling to the $A$ exciton situated at the $K$ point of the Brillouin zone. Owing to the structural relationship between strong intralayer bonding in two dimensions and weak interlayer interaction in the third dimension, like many properties of layered TMD systems, the wavefunctions of the excitons are expected to be very anisotropic. A recent work on the orientation of luminescent excitons in FL-MoS$_2$, isostructural to FL-WS$_2$, reveals them being confined entirely in-plane without significant expansion in the stacking direction of individual layers \cite{Schuller2013}. This was further substantiated by density-functional theory (DFT) calculations in the local-density approximation (LDA) that explicitely showed the wavefunction of the $A$ exciton in multilayer MoS$_2$ to be spread out over a large area in two dimensions but with negligible density in neighbouring layers \cite{Molina2013}. For the resonant Raman process discussed here this means that even for a layer number of more than one, the phonon couples to an $A$ exciton localized primarily in one of the layers. If the $N$ layers in FL-WS$_2$ are to be treated approximately as $N$ individual monolayers for the specific case of the $A$ excitonic resonant Raman process, there are $N$ Raman allowed Raman modes to be expected in the region of the monolayer A$'_1$ Raman mode with similar Raman intensities. They are still split in frequency due to interlayer interaction. Here, all $N$ components are identified in the Raman measurements presented in this article, but the infrared-active components - speaking from the $N$-layer symmetry point of view - are always weaker in intensity than the Raman active components. Thus we conclude that Raman selection rules are at least weakened but not completely broken. This is supported by the fact that the Raman active components are gaining in intensity relative to the main Raman peak but still appear as shoulders rather than as individual peaks.\\
Far from the $A$ excitonic resonance, the few-layer WS$_2$ can not be treated as $N$ individual monolayers and the Raman selection rules following from the few-layer symmetry strictly apply. This is also in agreement with recent findings on newly observed Raman modes in FL-MoS$_2$ in resonance with the $C$ exciton \cite{Scheuschner2015}. 

\section*{Conclusion}

In summary, we have shown experimentally as well as theoretically that the out-of-plane A$'_1$ mode of the WS$_2$ monolayer splits up in the few layer regime into \textit{N} components for \textit{N} layers. Despite the fact that only $\frac{N}{2}$ of them for even number of layers and $\frac{N+1}{2}$ for odd number of layers are Raman active, the full set of phonon modes is observed when the laser excitation energy is close to the $A$ excitonic transition energy. A possible explanation for this unusual behavior is presented by taking into account the in-plane orientation of the $A$ exciton wavefunction involved in the resonant Raman scattering process. The stiffening of the main out-of-plane phonon mode with \textit{N} is also followed by all other components successively added with increasing number of layers. By resonant Raman scattering measurements one can conclusively identify the number of layers in a specific sample by simply counting the number of components of the out-of-plane \textit{A} mode. The detailed analysis of the evolvement of the A$'_1$ mode of monolayer WS$_2$ presented here should in principle be applicable to (i) all other Raman modes of (ii) all layered materials in the few-layer regime.

\section*{Acknowledgements}

This work was supported by the Deutsche Forschungsgemeinschaft (DFG) in the Priority Programme SPP 1459 ``Graphene'' and by the European Research 
Council (ERC) through grant number 259286. We thank Lukas Madauß for his support with sample preparation.




\bibliographystyle{aip}

\end{document}